\documentclass[
groupedaddress,
notitlepage
]
{revtex4-1}
\RequirePackage{amsmath,amssymb,bm,wasysym}
\RequirePackage{graphicx}
\RequirePackage{hyperref}
\usepackage{siunitx}
\usepackage{tikz}
\usepackage{subcaption}
\usepackage{blkarray}
\usepackage[utf8]{inputenc}
\hypersetup{
breaklinks={true},
citecolor={blue},
colorlinks={true},
linkcolor={blue},
pdfcreator={\LaTeX and\flqq hyperref\frqq},
pdffitwindow={true},
pdfmenubar={true},
pdfpagelayout={OneColumn},
pdfstartview={FitH},
pdftoolbar={true},
plainpages={false},
pdfpagemode={UseThumbs},
}
\usepackage{braket}
\DeclareMathAlphabet{\mathpzc}{OT1}{pzc}{m}{it}

\newcommand{\minus}{\scalebox{0.45}[1.0]{$-$}}

\begin{document}
\pagestyle{plain}
\title{Relaxation of the Ising spin system coupled to a bosonic bath and the time dependent mean field equation}
\author{M\'at\'e Tibor Veszeli}
\affiliation{Institute of Physics, E\"{o}tv\"{o}s University, 1518 Budapest, Hungary}
\author{G\'abor Vattay}
\affiliation{Institute of Physics, E\"{o}tv\"{o}s University, 1518 Budapest, Hungary}
\pacs{
}
\begin{abstract}

The Ising model doesn't have a strictly defined dynamics, only a spectrum.
There are different ways to equip it with a time dependence e.g. the Glauber or
the Kawasaki dynamics, which are both stochastic, but it means there is a master equation which can also describes their dynamics.
We present a Gluber-type master equation derived from the Redfield equation, where the spin system is coupled to a bosonic bath. We derive a time dependent mean field equation which describes the relaxation of the spin system at finite temperature.
Using the fully connected, uniform Ising model the relaxation time will be studied, and the critical behaviour around the critical temperature. The master equation shows the finite size effects, and the mean field equation the thermodynamic limit.

\end{abstract}
\maketitle

\section{Introduction}
\label{sec:int}
Spin models are versatile, because they are simple, yet
able to demonstrate fundamental phenomenons, like phase transition
\cite{Erns-1925, PhysRev.65.117, Baxter-stat_mecha}.  Many complex physical
models can be reduced to a simple Ising or Heisenberg model, like electron and
nuclear spins \cite{solyom2007fundamentals}, and even social situations
\cite{mezard1987spin}. It is also important in modern applied physics since one
brach of adiabatic quantum computers - like the D-Wave system
\cite{harris2018phase} - are based on finding the global minimum of an
artificial spin system \cite{farhi2000quantum, roland2002quantum}. 

The Ising model is defined via its energy or in the quantum case, where it is often called Heisenberg model, via its Hamiltonian operator. The former do not have a natural dynamics, and although the latter has one, i.e. the Schrödinger or the Heisenberg equation, it is not always what we want. For example if we want our system to converge to the Boltzmann distribution, then  the Schrödinger equation is not enough.

To describe such a system we must use the tools of open quantum systems
\cite{breuer2002theory, schaller2014open} like the Redfield
\cite{redfield1965theory} and the Lindblad equation
\cite{lindblad1976generators}.  These equations have countless
applications in quantum biology \cite{rebentrost2009environment,
guerreschi2012persistent}, quantum optics \cite{breuer2002theory}, cold atomic gases \cite{sieberer2016keldysh},
chemical physics \cite{oppenheim1977stochastic} and
besides it is also relevant in quantum computing \cite{vacchini2010exact, davies1976quantum, leggett1987dynamics}.

Quantum dissipation and relaxation of spin systems in a bosonic bath and in magnetic field have been investigated by many authors.
\cite{albash2015decoherence, takada2016critical, cugliandolo2002dissipative, Sinha2013}. The interaction between an adiabatic computer and its enviroment is meant to be small, so the weak coupling Lindblad
equation will be used, but of course there are improved methods to describe open quantum systems,
like slippage initial condition \cite{suarez1992memory, gaspard1999slippage}, Nakajima-Zwanzig equation \cite{nakajima1958quantum, zwanzig1960ensemble}
or the polaron transformation \cite{wang2015nonequilibrium}.

The structure of this paper is the following.
In section \ref{sec:master_eq} we present a Glauber-type master equation based on the Redfield equation. In section \ref{sec:temp_eigvals} we investigate the temperature dependence of the eigenvalues of the transition matrix, because they contain relevant informations on the time scales of the system, e.g. the relaxation time. We give an upper bound to the smallest nonzero eigenvalue, then in section \ref{sec:uniform_eig_values} the dynamics of the uniform, fully connected Ising model is investigated, and we show that the relaxation time diverges in the thermodynamic limit as the temperature approaches the critical temperature.
In section \ref{sec:time_dep_mf} a time dependent mean field equation is derived from the master equation, which will be tested in section \ref{sec:time_dep_mf_uni_ising} using the uniform Ising model.

\section{Master equation of quantum Ising system}
\label{sec:master_eq}
\begin{figure}
 \begin{subfigure}{0.45\textwidth}
  \includegraphics[width=\textwidth]{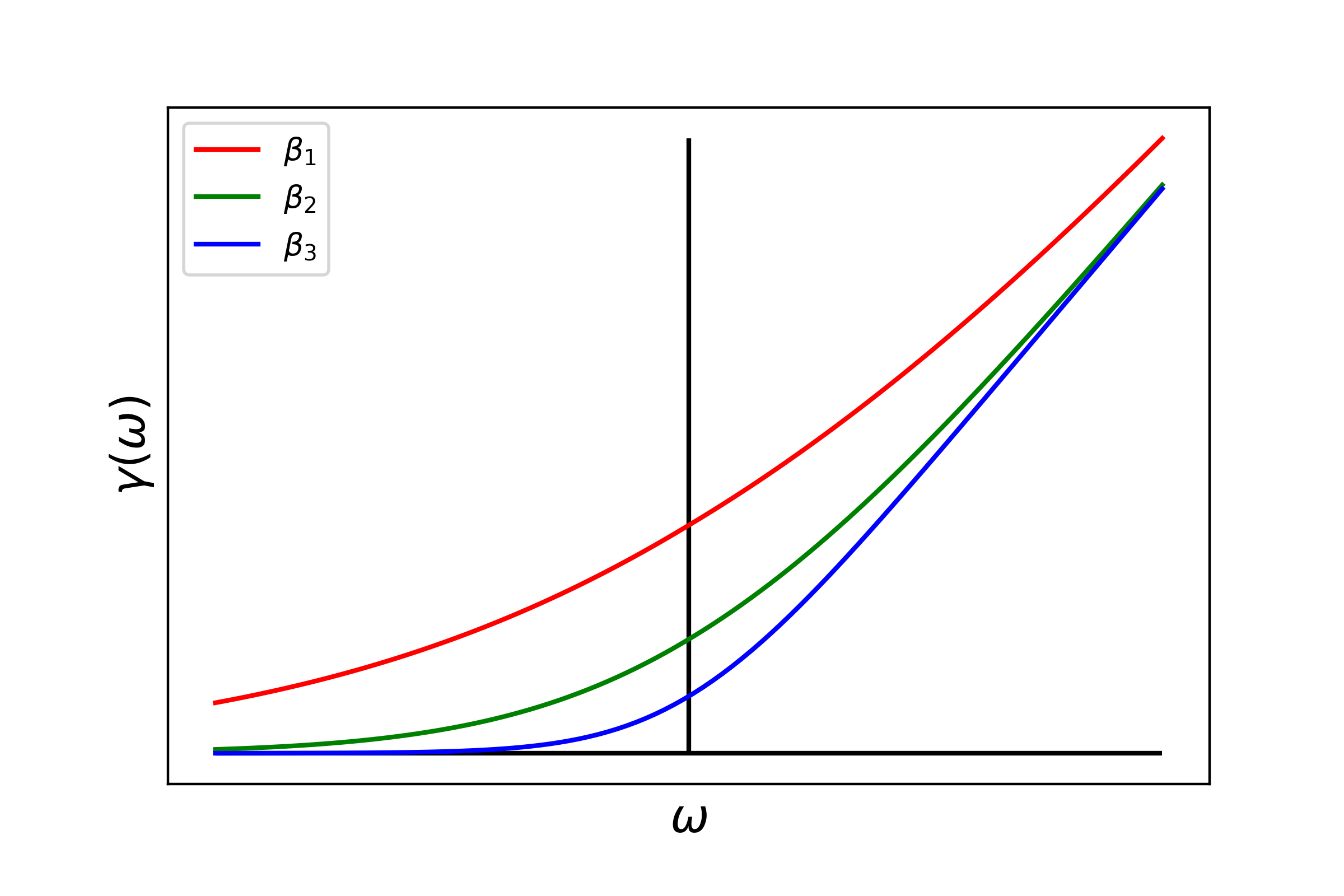}
  \caption{Ohmic case}
 \end{subfigure}
 \begin{subfigure}{0.45\textwidth}
  \includegraphics[width=\textwidth]{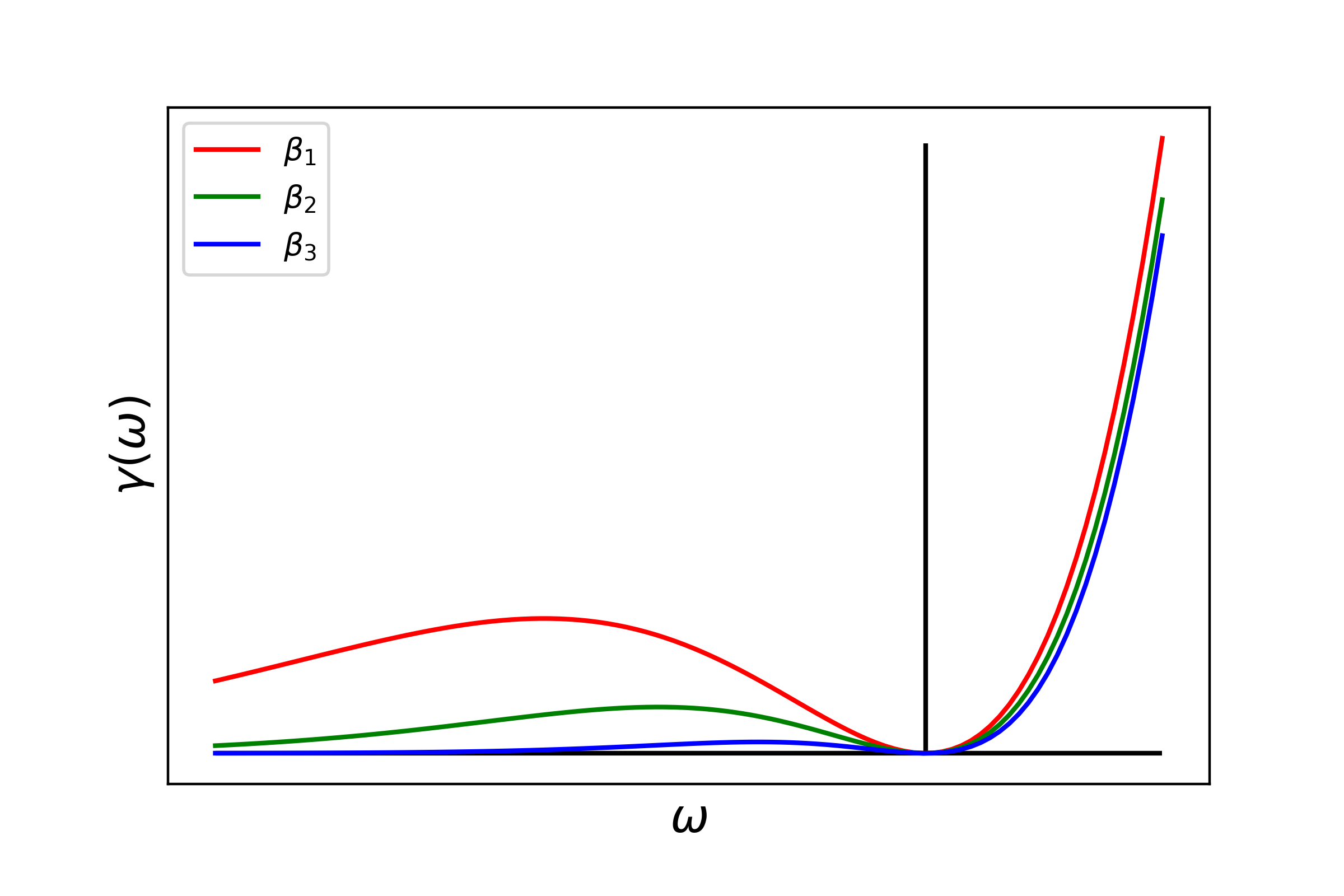}
  \caption{Super-Ohmic case}
 \end{subfigure}
 \caption{\textbf{Fourier transform of the bath correlation function}\\
  $\beta_1 < \beta_2 < \beta_3 $}
 \label{fig:bath_corr}
\end{figure}

In general if a system is connected to a bath, than its Hamiltonian operator is
\begin{equation}
 H_\text{tot} = H + H_\text{B} + H_\text{I},
\end{equation}
where $H$ acts only on the system of interest, $H_\text{B}$ only on the bath,
and $H_\text{I}$ is the interaction between the two subsystems, and it
can be written as $H_\text{I} = \sum_\alpha A_\alpha \otimes B$, where $A_\alpha$ and
$B_\alpha$ are system and bath operators respectively. The dynamics of the total
system is described by the von Neumann equation.
\begin{equation}
 \rho_\text{tot} = -i [H_\text{tot}, \rho_\text{tot}]
\end{equation}
If the interaction between the
system and the bath is small, than after the Born and the Markov approximation an effective equation can be derived to the
density matrix of the system of interest 
($\rho := \mathrm{Tr}_\text{B} \rho_\text{tot}$).
\begin{equation}
 \dot{\rho}(t) + i \left[ H, \rho(t) \right] =
 \sum_\alpha \left( A_\alpha  \rho(t) T_\alpha^\dag 
 - A_\alpha T_\alpha \rho(t) + \text{h.c.}
 \right),
\end{equation}
where 
$T_\alpha=\sum_\beta \int_0^\infty \mathrm{d}t 
C_{\alpha \beta}(t) A_\beta^\text{I}(-t)$,
$A_\beta^\text{I}(t)$ is in interaction picture,
$C_{\alpha \beta}(t) = \langle B_\alpha^\text{I}(t) B_\beta \rangle_\text{B}$ 
is the bath correlation function,
and h.c means hermitian conjugate.
This is the Redfield equation in weak-coupling limit \cite{redfield1965theory}.
After the so called secular or rotating wave approximation one can get to equation
\begin{equation}
\begin{aligned}
 \dot{\rho} + i \left[  H+H_\text{LS}, 
 \rho  \right]
 = \sum_{\alpha \beta} \sum_{\omega}
 & \gamma_{\alpha \beta}(\omega)
 \big( A_\beta(\omega) \rho A_\alpha^\dag(\omega)
 -\dfrac{1}{2} \left\{  A_\alpha^\dag(\omega) A_\beta(\omega),
  \rho \right\} \big),
\end{aligned}
\end{equation}
where $ A_\alpha(\omega) = 
\sum_{ i j } | i \rangle \langle i | A_\alpha | j \rangle \langle j |
\delta_{ \omega, \varepsilon_j - \varepsilon_i }$ and $|i\rangle$ is the
eigenvector of $H$ with eigenvalue $\varepsilon_i$ \cite{breuer2002theory}.
$H_\text{LS}$ is the Lamb shift Hamiltonian, which is usually small, so we will neglect it, and $\gamma_{\alpha \beta}(\omega)$ is the Fourier transform of the bath correlation function.
\begin{equation}
 \gamma_{\alpha \beta}(\omega) 
 := \int_{\minus \infty}^\infty \text{d}t e^{ i \omega t }
  \langle B^\dag_\alpha(t) B_\beta(0) \rangle_\text{B}
\end{equation}
There are two common bosonic bathes: the bath of phonons and the bath of
photons.
For phonons
$ \gamma_\text{ohm}(\omega) \sim e^{ -\frac{ |\omega| }{ \omega_c } }
 \dfrac{ \omega }{ 1 - e^{ \minus \beta \omega } }$, which is called
Ohmic case and for photons
$\gamma_\text{sup}(\omega) \sim
 \dfrac{ \omega^3 }{ 1 - e^{ \minus \beta \omega } }$, which is a super-Ohmic case.
 The frequency $\omega_c$
is the cutoff frequency. If we assume, that $ \omega_c$ is large compared
to the energy distances of the system, than
 $e^{ -\frac{ \omega }{ |\omega_\text{c}| } } \approx 1$.
Figure \ref{fig:bath_corr} shows the main features of the two $\gamma$ functions.
The main difference is that $\gamma_\text{ohm}$ is strictly increasing and
$\gamma_\text{ohm}(\omega=0) \sim k_\text{B}T$, but $\gamma_\text{sup}$ is
non-monotonic, and $\gamma_\text{sup} = 0$.
In the easiest case $\gamma_{\alpha \beta} \propto \delta_{\alpha \beta}$.

The advantage of the weak-coupling limit is that a master equation can be derived
to the diagonal elements of $\rho$.
\begin{equation}
\label{eq:master_eq}
 \dot P_i = \sum_j M_{i j} P_j \equiv \sum_j W_{i j} P_j - \sum_j W_{j i} P_i,
\end{equation}
where $P_i = \rho_{i i}$, $ W_{i j} = \sum_{\alpha \beta} 
\gamma_{\alpha \beta}\left( \omega_{j i}  \right)
(A_\alpha)_{j i} (A_\beta)_{i j}$ and $\omega_{j i} = 
(\varepsilon_j - \varepsilon_i)$.

The system converges to the Boltzmann
distribution if $W_{i j}$ satisfies the detailed balance condition i.e.
$W_{i j} = W_{j i} \exp(- \beta (\varepsilon_i - \varepsilon_j))$. Both the
Ohmic and the super-Ohmic bath satisfy it, because
\begin{equation}
\label{eq:gamma_detailed}
 \gamma( \minus \omega) = \mathrm{e}^{ \minus \beta \omega } \gamma(\omega)
\end{equation}

If the system of interest is the Ising model, then the Hamiltonian is
\begin{equation}
 H = -\sum_{ \underset{(i < j)}{i j} } J_{i j} \sigma_i^z \sigma_j^z -
  \sum_i h_i \sigma_i^z,
\end{equation}
where $\sigma_i^z$ is the Pauli z-matrix and the corresponding eigenvectors are
\begin{equation}
    | \underline{S} \rangle \equiv | S_1, S_2, \dots, S_N \rangle \qquad S_i \in \{  \pm 1 \}
\end{equation}
with eigenenergies
\begin{equation}
 E_{\underline{S}} =- \sum_{ \underset{(i < j)}{i j} } J_{i j}  S_i S_j
  - \sum_i h_i \sigma_i^z
\end{equation}
The easiest way to couple the system to the bath is via a Pauli matrix i.e.
$A_\alpha \rightarrow \sigma_i^x$. 
Using $\sigma^z$ in the interacion instead of
$\sigma^x$ would not give any relevant dynamics, since the system and the
interaction Hamiltonians would commute.
The peculiarity of this system is that the populations
decouple even without the secular approximation.

The $\sigma_i^x$ operator acting on $|\underline{S}\rangle$ only flips the $i$th spin, so the $W_{\underline{S}\, \underline{S}'}$ matrix element
is
\begin{equation}
\begin{aligned}
\label{eq:W_spin_master}
  W_{\underline{S}\, \underline{S}'} &= 
   \sum_i \gamma( \omega_{\underline{S}' \underline{S} } )
   (\sigma_i^x)_{\underline{S}' \underline{S}} (\sigma_i^x)_{\underline{S}\, \underline{S}'}
    &=
  \left\{\begin{matrix}
  \gamma( \omega_{\underline{S}' \underline{S} } ) 
   & | & \text{ if the Hamming distance between } \\ 
 & & \text{ $\underline{S}$ and $\underline{S}'$ is 1 }\\
 0 & | & \text{otherwise.}
\end{matrix}\right.
\end{aligned}
\end{equation}
With Eq. (\ref{eq:master_eq}) and (\ref{eq:W_spin_master}) we have a dynamics
for the Ising model.
\begin{equation}
\label{eq:master_spin_eq}
 \dot{P}_{\underline{S}} = \sum_{\underline{S}'}
  M_{\underline{S}\, \underline{S}'} P_{\underline{S}'} ,
\end{equation}
where $M_{\underline{S}\, \underline{S}'} = W_{\underline{S}\, \underline{S}'} - \delta_{\underline{S}\, \underline{S}'} \sum_{\underline{S}''} W_{\underline{S}'' \underline{S}'}$ is the transition matrix.
This matrix is temperature dependent, and it has at least one zero eigenvalue, which is the eigenvalue of the equilibrium distribution:
\begin{equation}
 P_{\underline{S}}^\text{eq} = \frac{ \mathrm{e}^{ -\beta E_{\underline{S}} } }{ Z }
\end{equation}
For constant temperature the general solution of (\ref{eq:master_spin_eq}) is
\begin{equation}
P_{\underline{S}}(t) = \sum_{ \underline{S}' } \sum_{\mu}
 \mathrm{e}^{ - \lambda_\mu t }
 P_{\mu, \underline{S}}^\text{R}
 P_{\mu, \underline{S}'}^\text{L}
 P_{\underline{S}}(t=0),
\end{equation}
 where $P_{\mu}^\text{R}$s are the right, and $P_{\mu}^\text{R}$s are the left eigenvectors of $M$ with $\minus \lambda_\mu$ eigenvalues. 
All the $\lambda_\mu$s are nonnegative. If the system is ergodic, then there is only one zero eigenvalue, and the other $\lambda$s are positive. Let the smallest positive be $\lambda_\text{min}$ and the largest be $\lambda_\text{max}$. The relaxation time is $t_\text{r} = 1/\lambda_\text{min}$. This is the time scale in which all but the equilibrium mode dies out. The other relevant time scale is $1/\lambda_\text{max}$, which is the characteristic time of the fastest mode. 
If for example this spin system is a quantum computer, then the fastest mode is the more important, because if the computation is slower than this time scale, then the enviroment isn't neglectable.
In other words $\lambda_\text{min}$ is important if we want the system to relax thermally, and $\lambda_\text{max}$ is important if we want to avoid any thermal influence.

\section{Temperature dependence of the eigenvalues}
\label{sec:temp_eigvals}
Both the smallest and the largest eigenvalue carry relevant information, and since $M(\beta)$ is temperature dependent $\lambda_\text{min}(\beta)$ and $\lambda_\text{max}(\beta)$  are too.

At high temperature we can determine the temperature dependence of all $\lambda$s by simply Taylor expanding $\gamma(\omega; \beta)$ for small
$\beta$.
\begin{equation}
 \gamma(\omega; \beta) = \eta \dfrac{ \omega^\alpha }
  { 1 - e^{ -\beta \omega } }  
  \approx \eta \dfrac{ \omega^{\alpha - 1} }{ \beta },
\end{equation}
where $\alpha=1$ in the Ohmic, and $\alpha=3$ in the super-Ohmic case. The
transition matrix inherits this temperature dependence:
$M_{\underline{S}\, \underline{S}'}(\beta) \sim \beta^{\minus 1}$, and hence
$\lambda \sim \beta^{\minus 1}$.

In spite of the high temperature limit, where the elements of the dynamical
matrix $M$ diverges, in the low temperature limit they
converge.
\begin{equation}
\begin{aligned}
 \lim_{\beta \rightarrow \infty} \gamma(\omega ; \beta) = &
   \left\{\begin{matrix}
     0  & | & \omega \leq 0 \\ 
     \eta \omega^\alpha & | & \omega > 0
  \end{matrix}\right. .
\end{aligned}
\end{equation}
It means all the eigenvalues also converge.  As a consequence we can't slow
down arbitrary all the modes by reducing the temperature. We have an upper
limit in time for the quantum computing. Of course this calculation is valid
only for a time independent system, but the main features apply to more general
cases.

Without external magnetic field ($\underline{h}=0$) at zero temperature the equilibrium Boltzmann distribution prefers only the two spin configurations with the lowest energies:
\begin{equation}
 P_{\underline{S}}^\text{eq}(T=0) = \frac{ 1 }{ 2 }
 ( \delta_{\underline{S},\underline{S}_\text{g}} 
 + \delta_{\underline{S},\minus \underline{S}_\text{g}}
),
\end{equation}
where $\underline{S}_\text{g}$ and $\minus \underline{S}_\text{g}$ are the ground states.
At zero temperature there is one more eigenvector with zero eigenvalue:
\begin{equation}
 P_{\text{min}, \underline{S}}^\text{R} = 
 \frac{ 1 }{ 2 }  ( \delta_{\underline{S},\underline{S}_\text{g}} 
 - \delta_{\underline{S},\minus \underline{S}_\text{g}}
)
\end{equation}
The question is how $\lambda_\text{min}(\beta)$ behaves at low temperature. We can give an upper bound.
First let us introduce the following symmetric matrix:
\begin{equation}
 \tilde{M}_{ \underline{S}\, \underline{S}' } = 
   M_{ \underline{S}\, \underline{S}' }
 \sqrt{ \dfrac{ P_{\underline{S}'}^\text{eq} }{ P_{\underline{S}}^\text{eq}} }
 \equiv M_{ \underline{S}\, \underline{S}' }
 e^{ -\beta \frac{ E_{{\underline{S}'}} - E_{\underline{S}} }{ 2 } }
\end{equation}
This transformation doesn't affect the eigenvalues, and the eigenvectors
transform like
\begin{equation}
 \tilde{P}_{\mu \underline{S}} = 
  \dfrac{ P_{\mu, \underline{S}}^\text{R} }{ \sqrt{P_{\underline{S}}^\text{eq}} }.
\end{equation}
Since $\tilde{M}$ is symmetric its right and left eigenvectors are the same, and now the variational method applies to it:
\begin{equation}
\label{eq:lambda_min_var}
 \lambda_\text{min} \leq
   - \sum_{ \underline{S}, \underline{S}' } \tilde{\Pi}_{\underline{S}} 
  \tilde{M}_{\underline{S}\, \underline{S}'} \tilde{\Pi}_{\underline{S}'},
\end{equation}
where $\tilde{\Pi}$ is an arbitrary vector with $\sum_{\underline{S}} \tilde{\Pi}_{\underline{S}}^2 = 1$, and it must be perpendicular to the equilibrium vector ( $\tilde{P}_{\underline{S}}^\text{eq} \equiv
  \sqrt{ P_{\underline{S}}^\text{eq} } $ ), because $\lambda_{\text{min}}$ is the second smallest eigenvalue of $\tilde{M}$.
Let $\tilde{\Pi} = 
 \frac{ 1 }{ \sqrt{2} } 
 ( \delta_{\underline{S}\, \underline{S}_\text{g} }
 -\delta_{\underline{S}\, -\underline{S}_\text{g} } )  $.
Then accordig to (\ref{eq:lambda_min_var})

\begin{equation}
\begin{aligned}
\label{eq:var_princ2}
 \lambda_\text{min} &\leq
 - \dfrac{ 1 }{ 2 } (
   \tilde{M}_{ \mathbf{S}_\text{g},  \mathbf{S}_\text{g}} -
   \tilde{M}_{ \mathbf{S}_\text{g}, \minus \mathbf{S}_\text{g}} -
   \tilde{M}_{ \minus \mathbf{S}_\text{g},  \mathbf{S}_\text{g}} +
   \tilde{M}_{ \minus \mathbf{S}_\text{g}, \minus \mathbf{S}_\text{g}} 
 ) \\
 &= - \tilde{M}_{ \mathbf{S}_\text{g},  \mathbf{S}_\text{g}}
 = - M_{ \mathbf{S}_\text{g},  \mathbf{S}_\text{g}} = 
 \sum_\mathbf{S} W_{\mathbf{S}, \mathbf{S}_\text{g}}\\
 &= \sum_{ \underset{
   d(\mathbf{S},\mathbf{S}_\text{g})=1}{ \mathbf{S}} }
   \gamma( \omega_{\mathbf{S}_\text{g}, \mathbf{S}}; \beta ),
\end{aligned}
\end{equation}
where $d(\underline{S},\underline{S}_\text{g})$  is the Hamming distance,
and the $\underline{S} \mapsto \minus \underline{S}$ symmetry was used. In the bosonic bath
\begin{equation}
 \lambda_{\text{min}} ( \beta ) \leq 
  \sum_{ \underset{ d(\underline{S},\underline{S}_\text{g}) = 1 }
   {\underline{S} } }
 \eta \big(  \Delta E_{\underline{S}} \big)^\alpha
 \frac{ 1 }{ \mathrm{e}^{ \beta \Delta E_{\underline{S}} } - 1 },
\end{equation}
where $\Delta E_{\underline{S}} := E_{\underline{S}} -
 E_{\underline{S}_\text{g}} > 0$.
At low temperature this is the sum of some $\mathrm{e}^{ - \beta \Delta E_{\underline{S}} }$ functions, so $\lambda_{\text{min}}( \beta )$ can be estimated from above with an exponential function.

\begin{figure}
\centering
\captionsetup{justification=centering}
 \begin{subfigure}{0.45\textwidth}
  \includegraphics[width=\textwidth]{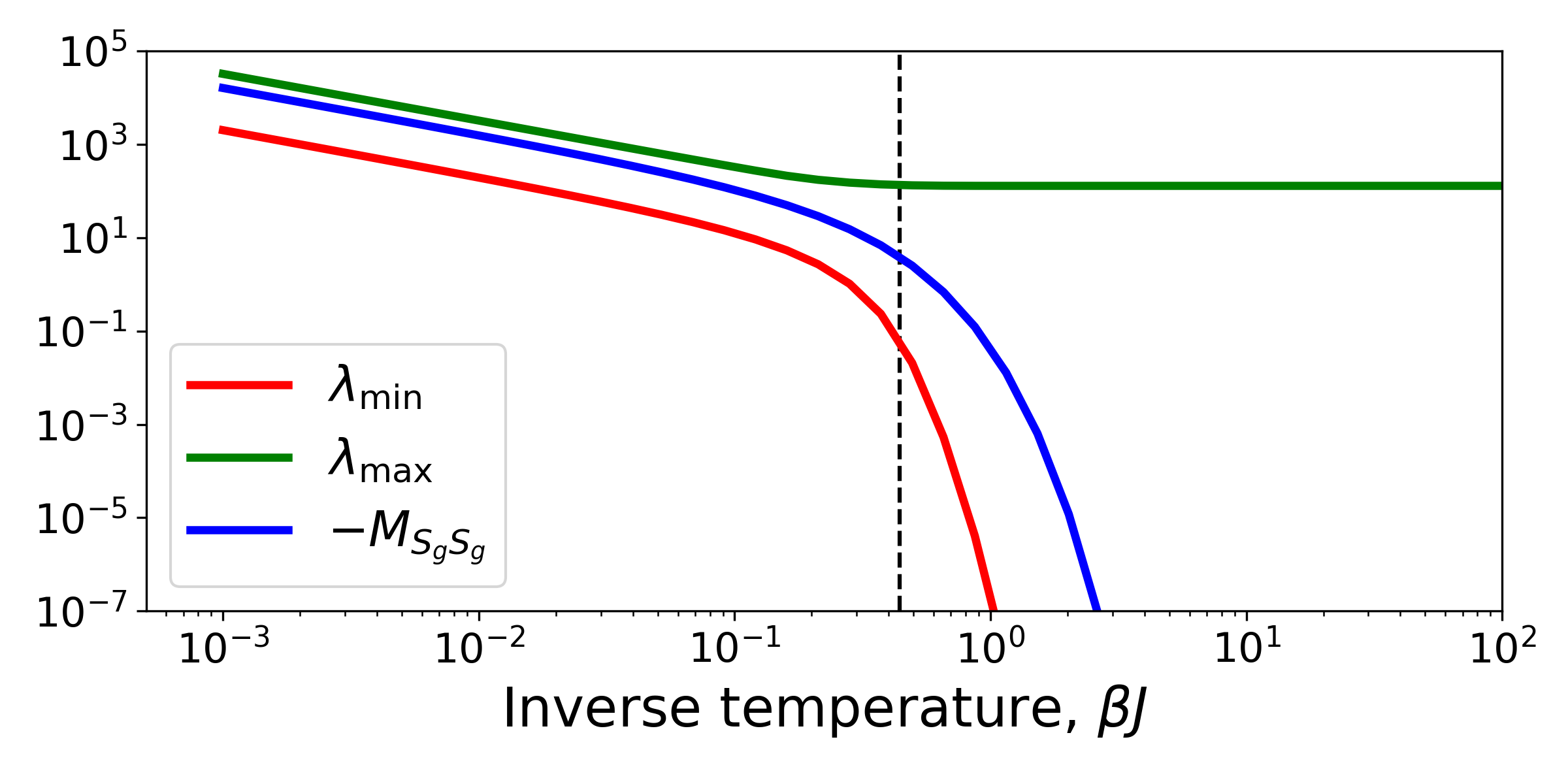}
 \end{subfigure}
 \begin{subfigure}{0.45\textwidth}
  \includegraphics[width=\textwidth]{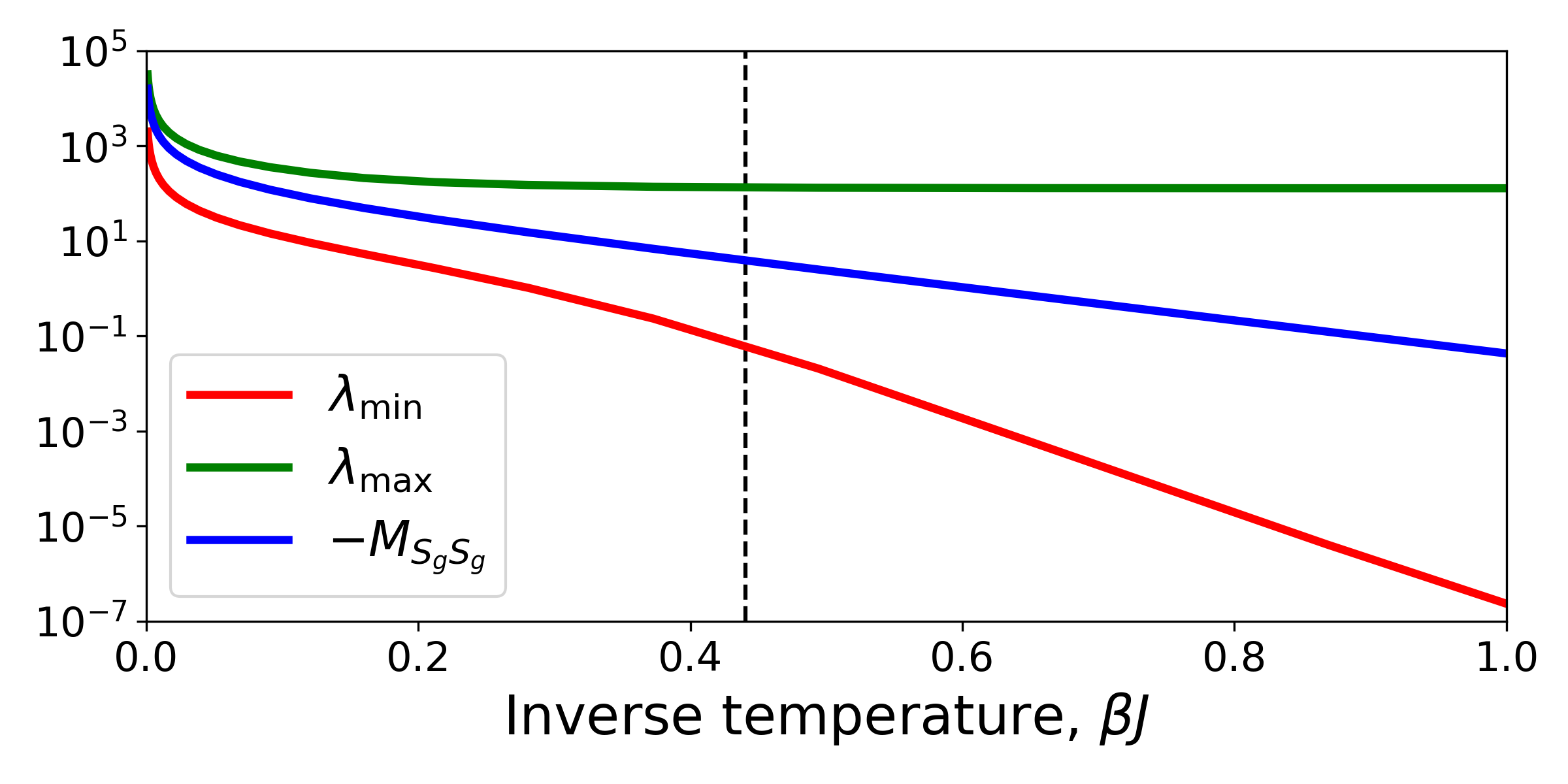}
 \end{subfigure}
 \caption{\textbf{Temperature dependence of $\lambda_\text{min}$, $\lambda_\text{max}$  and $M_{\underline{S}_\text{g}, \underline{S}_\text{g}}$ }\\
 2D, ferromagnetic, $4 \times 4$ Ising model, Ohmic bath, $J=1$, $\eta=1$
}
 \label{fig:lambda_temp}
\end{figure}

Figure \ref{fig:lambda_temp} shows $\lambda_{\text{min}}( \beta )$, $\lambda_{\text{max}}( \beta )$ and $\minus M_{\underline{S}_\text{g}, \underline{S}_\text{g}}$ (the upper bound) for a $4 \times 4$, ferromagnetic, 2D Ising model with Ohmic bath. The dashed vertical line marks the critical temperature ($\beta_\text{c} J = \frac{ \ln(1 + \sqrt{2}) }{ 2 } \approx 0.44$). The left figure is in log-log scale, where we can see, that at low temperature the eigenvalues has a $\beta^{\minus 1}$ temperature dependence, and $\lambda_\text{max}(\beta)$ converges, and the right figure with lin-log scale shows, that $\lambda_\text{min}(\beta)$ goes to zero exponentially.

\section{Eigenvalues of the uniform Ising model}
\label{sec:uniform_eig_values}
The $M$ matrices are $2^N \times 2^N$ large, therefore we can't see how the eigenvalues behave at the thermodinamic limit. However the uniform, fully connected Ising model is so symmetric, that an effective equation can be derived, which has the same relaxation time as the original equation.  

\begin{figure}
\centering
\captionsetup{justification=centering}
 \begin{subfigure}{0.45\textwidth}
  \includegraphics[width=\textwidth]{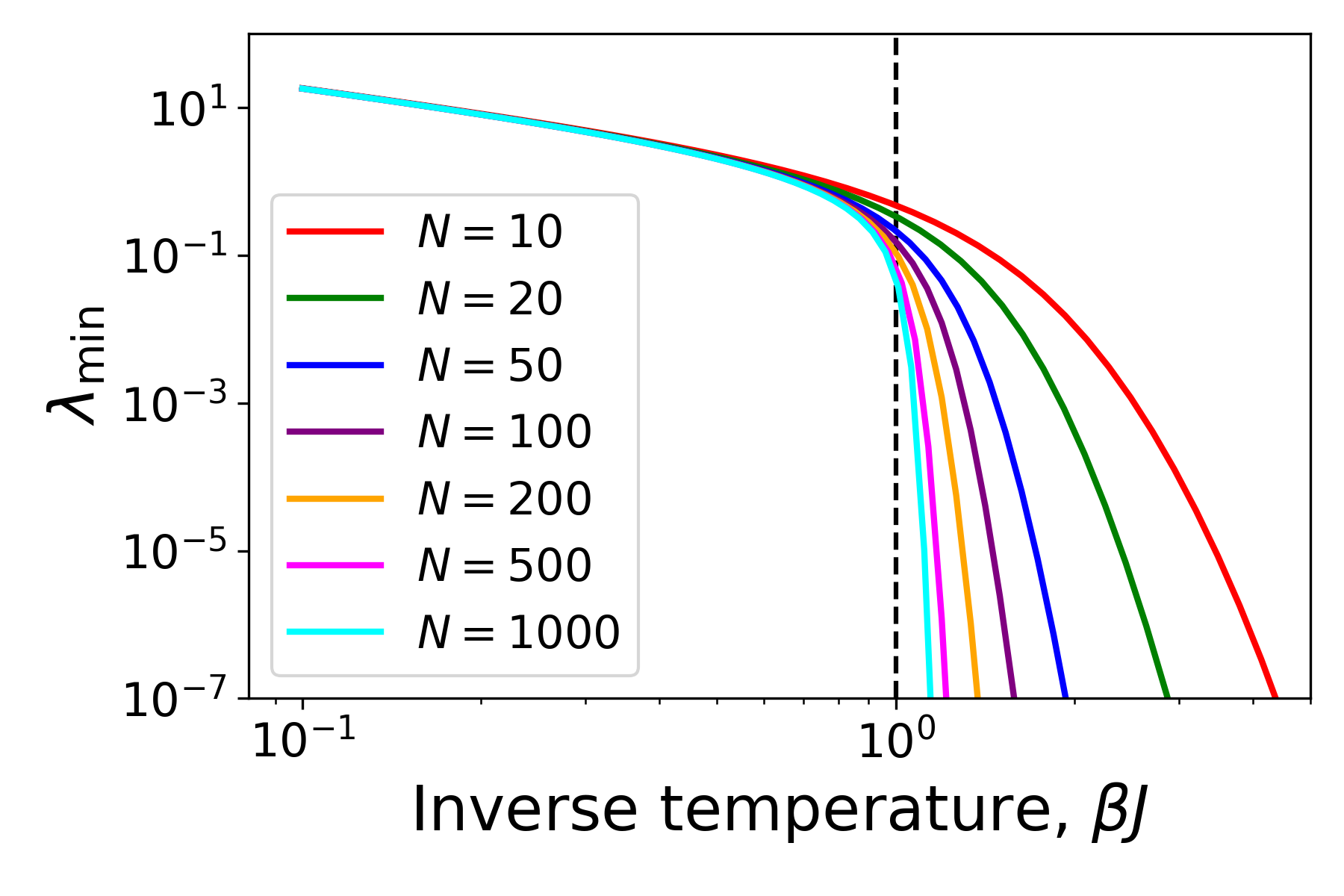}
  \caption{ \textbf{ Temperature dependence of $\lambda_{\text{min}}$ for different system sizes } }
  \label{fig:beta-lambda_large}
 \end{subfigure}
 \begin{subfigure}{0.45\textwidth}
  \includegraphics[width=\textwidth]{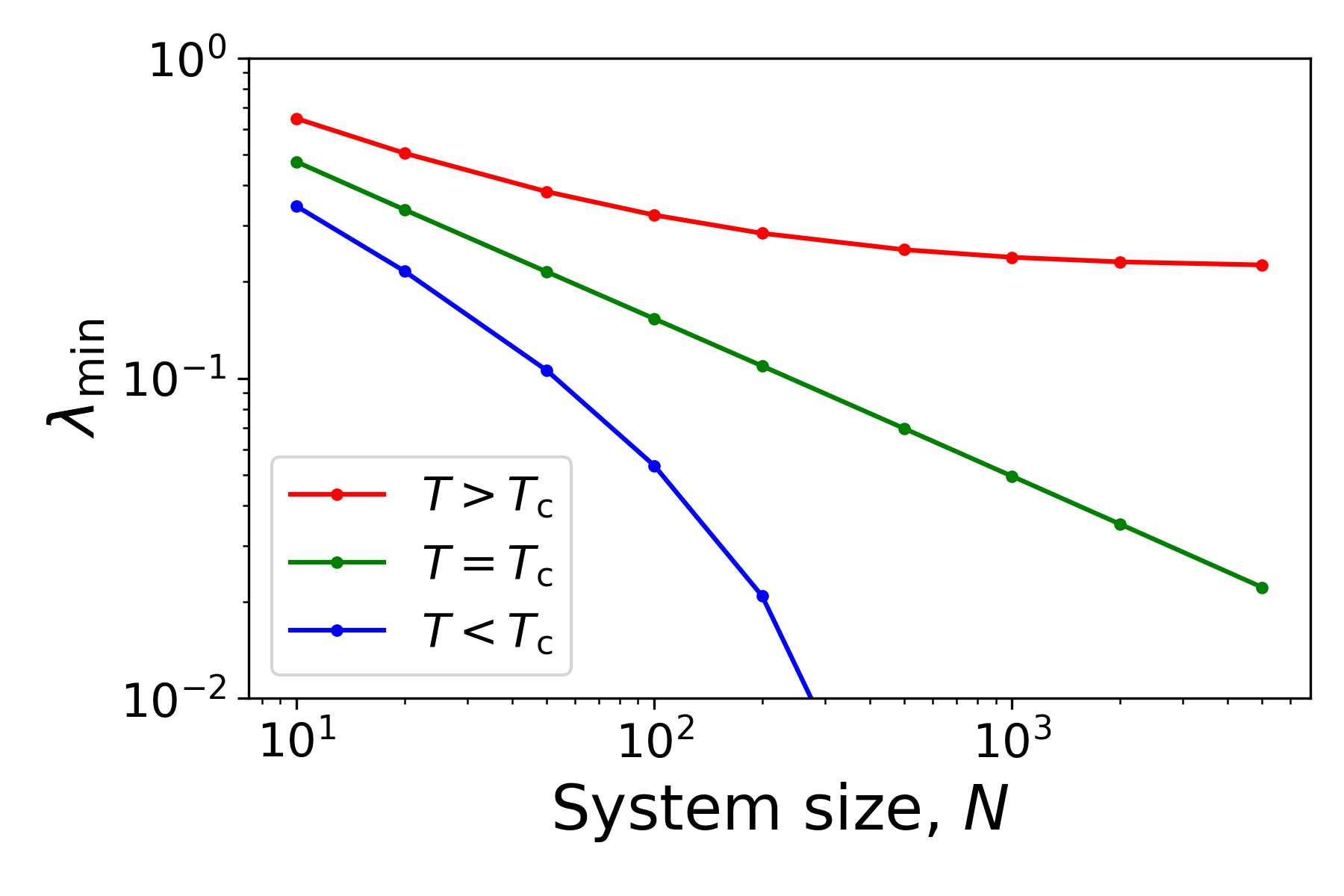}
  \caption{ \textbf{System size dependence of $\lambda_{\text{min}}$ below, above and at the critical temperature } }
  \label{fig:N-lambda_large}
 \end{subfigure}
 \caption{  
   \textbf{Fully connected, uniform Ising model, Smallest eigenvalue of $M^\text{red}$}\\
  $J=1$, $\eta=1$
  }
\end{figure}

The energy of the model is
\begin{equation}
 E_{\underline{S}} = - \frac{ J }{ N } \sum_{ \underset{(i > j)}{i,j=1} }^N S_i S_j.
\end{equation}
The $1/N$ factor is to keep the energy extensive and $J>0$. Given an $\underline{S}$ microstate, it consists of $N_\uparrow$ spins with $S_i=1$ and $N_\downarrow$ spins with $S_i=-1$. The number of spins is constant, i.e. $N_\uparrow + N_\downarrow = N = \text{fix}$. The energy of such a configuration is
\begin{equation}
\label{eq:uni_ising_energy}
 E_{\underline{S}} = - \frac{ J }{ N } \left[ 
  \frac{ N_\uparrow (N_\uparrow - 1) }{ 2 } 
  + \frac{ N_\downarrow (N_\downarrow - 1) }{ 2 }
  - N_\uparrow N_\downarrow
 \right],
\end{equation}
If $N$ is fixed, then the energy is the function of only $N_{\uparrow}$. The symmetry of the system is that we can perturb the spins any way, the energy and the $M$ matrix remains the same. If in the dynamic the initial condition also has this symmetry, then the $P_{\underline{S}}$ will inherit this property. The slowest mode propagates between the two  deepest valley of the energy landscape, which are the $\uparrow \uparrow \dots \uparrow$ and $\downarrow \downarrow \dots \downarrow$. Assume that initially $P_{\downarrow \downarrow \dots \downarrow}(t=0) = 1$, and we want to determine relaxation time, where
$P_{\downarrow \downarrow \dots \downarrow}(t_\text{r}) \approx P_{\uparrow \uparrow \dots \uparrow}(t_\text{r})$.
Since both the equations and the initial condition has the permutation symmetry all the probabilities, which has the same up spin has the same value, e.g. for 3 spins $P_{\uparrow \downarrow \downarrow}(t) = P_{\downarrow \uparrow \downarrow}(t) = P_{\downarrow \downarrow \uparrow}(t) \ \forall t$. The probability can only flow between spin configurations if the Hamming distance between them is 1.
Let us introduce the following probabilities:
\begin{equation}
 P_{N_\uparrow} = \sideset{}{'}\sum_{\underline{S}} P_{\underline{S}} = 
  \binom{N}{N_{\uparrow}} 
  P_{ \underbrace{\uparrow \dots \uparrow}_{N_{\uparrow}}
      \underbrace{\downarrow \dots \downarrow}_{N - N_{\uparrow}}  },
\end{equation}
where the prime denotes that only such configurations count where there are $N_{\uparrow}$ up spin. We can give a closed set of differential equations which only contain this new $P_{N_\uparrow}$ probabilities.
\begin{equation}
\begin{aligned}
\label{eq:reduced_master_uniform}
 \dot{P}_{N_{\uparrow}} = &
 \binom{N}{N_{\uparrow}} \Bigg[
  N_{\uparrow} W_{N_{\uparrow}, N_{\uparrow}-1} 
   \frac{ P_{N_\uparrow-1} }{ \binom{N}{N_{\uparrow}-1} }
 + (N-N_{\uparrow})  W_{N_{\uparrow}, N_{\uparrow}+1}
  \frac{ P_{N_\uparrow+1} }{ \binom{N}{N_{\uparrow}+1} }
  \Bigg] \\
  &- \left( 
     N_{\uparrow} W_{N_{\uparrow}-1, N_{\uparrow}} 
     + (N-N_{\uparrow}) W_{N_{\uparrow}+1, N_{\uparrow}}
    \right) P_{N_{\uparrow}}\\
  =&
   (N-N_{\uparrow}+1) W_{N_{\uparrow}, N_{\uparrow}-1} P_{N_\uparrow-1} +
    (N_{\uparrow}+1) W_{N_{\uparrow}, N_{\uparrow}+1} P_{N_\uparrow+1}\\
  &- \left( N_{\uparrow} W_{N_{\uparrow}-1, N_{\uparrow}} 
     + (N-N_{\uparrow}) W_{N_{\uparrow}+1, N_{\uparrow}}
    \right) P_{N_{\uparrow}},
\end{aligned}
\end{equation}
where $W_{N_{\uparrow}, N_{\uparrow}+1} = \gamma( E_{N_\uparrow + 1} - E_{N_\uparrow} ) = \gamma \left( -\frac{ 2 }{ N } ( 2N_{\uparrow} - N + 1 ) \right) $.
This master equation has only $N+1$ variables instead of $2^N$, thus easy to simulate for large systems.
A comparison between the quantum and the thermal simulated annealing of the fully connected Ising model was investigated by Wauters et al. using a similar reduced master equation \cite{wauters2017direct}.
Equation (\ref{eq:reduced_master_uniform}) has the form
\begin{equation}
\label{eq:reduced_master_uniform2}
 \dot{P}_{N_{\uparrow}} = \sum_{ N_{\uparrow}' = 0 }^N 
  M_{N_{\uparrow}, N_{\uparrow}'}^\text{red} P_{N_{\uparrow}'},
\end{equation}
and we want to determine the lowest (nonzero) eigenvalue of $M^\text{red}$, which is the same as the lowest (nonzero) eigenvalue of $M$. The matrix $M^\text{red}$ is sparse, because it is a tridiagonal matrix, i.e. only the main diagonal, the first diagonal below and above the main diagonal is nonzero. Figure \ref{fig:beta-lambda_large} shows the temperature dependence of $\lambda_{\text{min}}$ for different system sizes. As $N$ increases we can see, that around the critical temperature (which is $\beta_\text{c} J = 1$) the behaviour of the system changes. At figure \ref{fig:N-lambda_large} we can see it better, that above the critical temperature ($T>T_{\text{c}}$) for large $N$ values $\lambda_{\text{min}}$ converges, meaning for every system size there is a finite relaxation time. At the critical temperature $(T = T_{\text{c}})$, it follows a power law ($\lambda_{\text{min}} \propto N^{\minus 0.5} $). Below the critical temperature $(T < T_{\text{c}})$ $\lambda_{\text{min}}$ goes to 0 for large $N$, but doesn't follow a power law.
This behaviour is the famous critical slowing down phenomenon.

\begin{figure}
\centering
\captionsetup{justification=centering}
 \includegraphics[width=0.6\textwidth]{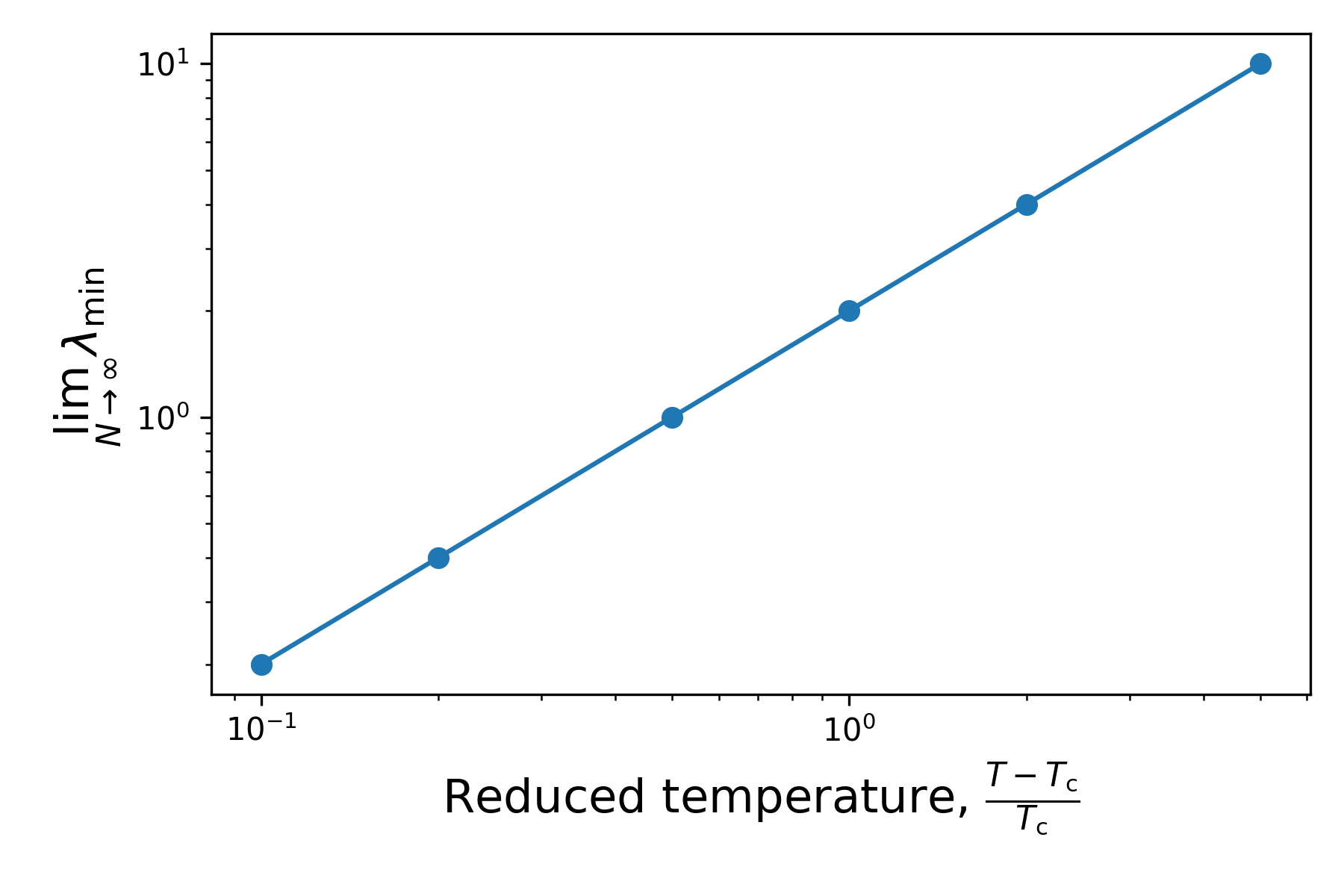}
 \caption{
  \textbf{Critical behaviour of the fully connected Ising model above $T_\text{c}$.}\\
 $J=1$, $\eta=1$
 }
 \label{fig:dynamic_crit_exponent}
\end{figure}

From the $N \rightarrow \infty$ thermodinamic limit we can determine the dynamical critical exponent. Figure \ref{fig:dynamic_crit_exponent} shows $\lambda_\text{min}(T, N \rightarrow \infty)$ as the function of the reduced temperature ($\frac{ T - T_{\text{c}} }{ T_{\text{c}} }$). This follows an easy power law, because $\lambda_\text{min} \propto T-T_{\text{c}}$. In the next section we will see that this result can be obtained from the mean field approximation.

\section{ Time dependent mean field equation }
\label{sec:time_dep_mf}

Since the primary interest is the magnetization ($m_i := \langle S_i \rangle$), we would like to derive a differential equation for it.
Using the definition of $m_i$ and the master equation we get
\begin{equation}
\label{eq:dot_m_W}
 \dot{m}_i = \sum_{\underline{S}} \dot{P}_{\underline{S}} S_i =
 \sum_{\underline{S}\, \underline{S}'} W_{\underline{S}\, \underline{S}'}
 P_{\underline{S}'} S_i
 - \sum_{\underline{S}\, \underline{S}'} W_{\underline{S}' \underline{S}}
 P_{\underline{S}} S_i'
 = \sum_{\underline{S}' \underline{S}} W_{\underline{S}' \underline{S}}
 P_{\underline{S}} (  S_i' - S_i )
\end{equation}
 The $W_{\underline{S}' \underline{S}}$ matrix component is nonzero if the Hamming distance between $\underline{S}'$ and $\underline{S}$ is one. Introducing
\begin{equation}
\begin{aligned}
 \Lambda_i(\underline{S},n)
 =\left\{\begin{matrix}
  S_i & | &  i \neq n \\ 
  \minus S_i & | &  i = n
\end{matrix}\right. 
\end{aligned}
\end{equation}
we can rewrite the double sum in (\ref{eq:dot_m_W}).
\begin{equation}
 \label{eq:dot_m_n}
 \dot{m}_i = \sum_{\underline{S}} \sum_{n=1}^N
 W_{ \underline{\Lambda}(\underline{S},n), \underline{S} } \,
 P_{\underline{S}} ( \Lambda_i(\underline{S},n) - S_i )
= \sum_{\underline{S}}
 W_{ \underline{\Lambda}(\underline{S},i), \underline{S} } \,
 P_{\underline{S}} ( - 2 S_i )
 = -2 \langle W_{ \underline{\Lambda}(\underline{S},i), \underline{S} }
    S_i \rangle
\end{equation}
In the second step the $(\Lambda_i(\underline{S},n) - S_i) = -2 S_i \delta_{i n}$ identity was used.
The nonzero elements of $W$ are the function of the energy difference:
\begin{equation}
 W_{ \underline{\Lambda}(\underline{S},i), \underline{S} } =
 \gamma( E_{\underline{S}} - E_{\underline{\Lambda}(\underline{S},i)} )
 = \gamma \Big( -2 \Big(\sum_j J_{i j} S_j + h_i\Big) S_i \Big) \equiv
  \gamma \Big( -2 \tilde{h}_i S_i \Big), 
\end{equation}
where $\tilde{h}_i = \sum_j J_{i j} S_j + h_i$, so this is still the function of the $\underline{S}$ random variable, but because $J_{i i} = 0$ it is not a function of $S_i$. Since $S_i$ can be only $1$ or $\minus 1$ the $\gamma(-2 \tilde{h}_i S_i)$ as a function of $S_i$ must have the
\begin{equation}
 \gamma(-2 \tilde{h}_i S_i ) \equiv
  \frac{ \gamma(-2 \tilde{h}_i ) + \gamma(2 \tilde{h}_i ) }{ 2 }
 + \frac{ \gamma(-2 \tilde{h}_i ) - \gamma(2 \tilde{h}_i ) }{ 2 } S_i
\end{equation}
form. Using (\ref{eq:gamma_detailed}) yields
\begin{equation}
 \gamma(-2 \tilde{h}_i S_i ) =
 \gamma(2 \tilde{h}_i ) \left[ 
  \frac{ \mathrm{e}^{ -2\beta \tilde{h}_i } + 1 }{ 2 }
 +\frac{ \mathrm{e}^{ -2\beta \tilde{h}_i } - 1 }{ 2 } S_i
 \right]
 = \gamma(2 \tilde{h}_i )
  \frac{ \mathrm{e}^{ -2\beta \tilde{h}_i } + 1 }{ 2 }
  \left[ 
  1 -  \tanh( \beta \tilde{h}_i ) S_i
 \right],
\end{equation}
 then substituting back to (\ref{eq:dot_m_n}) gives
\begin{equation}
\label{eq:callen_like}
 \dot{m}_i = 
 - \Big\langle \gamma(2 \tilde{h}_i )
  \left( \mathrm{e}^{ -2\beta \tilde{h}_i } + 1 \right)
 \left(  S_i - \tanh(\beta \tilde{h}_i) \right) \Big\rangle.
\end{equation}
Equation (\ref{eq:callen_like}) is similar to the Callen equation \cite{callen1963note, parisi1988statistical}
$(  
 \langle S_i \rangle = \langle \tanh( \beta \tilde{h}_i ) \rangle
)$, 
where the averaging is outside the hyperbolic function.
In order to get a closed equation to the expected values the average must move inside, and instead of the $S_i$ random variables their $m_i$ expected values must be written.
\begin{equation}
\label{eq:dot_m_beta}
 \dot{m}_i = 
 -\gamma\Big( 2 ( \Sigma_j J_{i j} m_j + h_i)  \Big)
  \left( 1 + \mathrm{e}^{ -2\beta ( \sum_j J_{i j} m_j + h_i) }  \right)
 \Big(  m_i - \tanh \big(\beta ( \Sigma_j J_{i j} m_j + h_i) \big) \Big)
\end{equation}
 The right-hand side contains the self-consistent equation from the equilibrium statistical physics, hence if the equation of state is satisfied, then $\dot{m}_i = 0$.

Equation (\ref{eq:dot_m_beta}) contains both the real time and the temperature of the bath. The temperature can be also time dependent, and in that case what we could get is a thermal annealing, but if the temperature is constant we can determine the relaxation time, and the dynamical critical exponent.
If 
$\underline{m}(t) = \underline{m}^\text{eq} + \delta \underline{m}(t)$,
where $\underline{m}^\text{eq}$ is the equilibrium solution and $\delta \underline{m}(t)$ is small, then the linearized equation of (\ref{eq:dot_m_beta}) is
\begin{equation}
 \delta \dot{m}_i = - b_i( \underline{m}^\text{eq} )
 \sum_j \left\{ \left( 
  \delta_{i j} -
   \frac{ \beta J_{i j} }{ \cosh^2( \beta \sum_k J_{i k} m_k^\text{eq} + h_i ) }
 \right) \delta m_j \right\},
\end{equation}
where 
$b_i( \underline{m}^\text{eq} ) = 
\gamma \big( 2 ( \Sigma_j J_{i j} m_j^\text{eq} + h_i)  \big)
  \left( 1 + \mathrm{e}^{ -2\beta ( \sum_j J_{i j} m_j^\text{eq} + h_i) }  \right)
 $.
Using the $\frac{ 1 }{ \cosh^2(x) } \equiv 1 - \tanh^2(x)$ identity, and the equation of state we get to
\begin{equation}
\label{eq:lin_dot_m}
 \delta \dot{m}_i = - b_i( \underline{m}^\text{eq} )
 \sum_j \left\{ \left( 
  \delta_{i j} -
   \beta J_{i j} \left( 1 - \left( m_j^\text{eq} \right)^2 \right)
 \right) \delta m_j \right\}.
\end{equation}
Equation (\ref{eq:lin_dot_m}) contains the inverse susceptibility of the mean field Ising model.
\begin{equation}
 \chi_{i j}^{\minus 1} := \frac{ \partial^2 F^\text{MFA} }{ \partial m_i \partial m_j  }
 = - J_{i j} + \frac{ T \delta_{i j} }{ 1-m_i^2 },
\end{equation}
where 
\begin{equation}
\begin{aligned}
\label{eq:mean_field_free_energy}
F^\text{MFA}(\underline{m}, \underline{h}, T) 
 = - \frac{ 1 }{ 2 } \sum_{i j} J_{i j} m_i m_j
- \sum_i h_i m_j
+ T
\sum_i \Bigg[ \dfrac{1+m_i}{2} \ln\left( \frac{1+m_i}{2} \right)
+\dfrac{1-m_i}{2} \ln\left( \frac{1-m_i}{2} \right)
\Bigg].
\end{aligned}
\end{equation}
Substituting the inverse susceptibility back into (\ref{eq:lin_dot_m}) yields
\begin{equation}
\label{eq:lin_dot_m2}
 \delta \dot{m}_i = - b_i( \underline{m}^\text{eq} ) \beta
 \left( 1 - \left( m_i^\text{eq} \right)^2 \right)
 \sum_j \chi_{ij}^{\minus 1} \delta m_j
 \equiv - \Gamma_i \sum_j \chi_{i j}^{\minus 1} \delta m_j.
\end{equation}
This is a well known equation in the theory of dynamical critical phenomena \cite{hohenberg1977theory}, but it is usually derived from the
$
\dot{\underline{m}} = - \Gamma \partial_{\underline{m}} F^\text{MFA}
$
phenomenological equation.
Now we can see, how it is related to a master equation and the spin-boson model.
If the system is symmetric in a sense, that all the spins behave the same, then (\ref{eq:lin_dot_m2}) simplifies to
\begin{equation}
 \delta \dot{m} = - \Gamma \chi^{\minus 1} \delta m,
\end{equation}
where $\chi^{\minus 1} = \sum_j \chi_{i j}^{\minus 1}$.

\section{Time dependent mean field equation for the uniform Ising model}
\label{sec:time_dep_mf_uni_ising}
As before in section \ref{sec:uniform_eig_values} the uniform Ising model will be studied, because in the equilibrium case in the thermodynamic limit it gives back the exact results.
According to (\ref{eq:mean_field_free_energy}) the mean field free energy is
\begin{equation}
 \frac{F^\text{MFA}(m,h,T)}{N} = -\frac{ 1 }{ 2 } J m^2 - h m +
  T \left( \frac{ 1+m }{ 2 }  \ln\left( \frac{ 1 + m }{ 2 } \right) +
  \frac{ 1-m }{ 2 }  \ln\left( \frac{ 1 - m }{ 2 } \right) \right),
\end{equation}
and the time dependent mean field equation is
\begin{equation}
\label{eq:uniform_dot_m}
 \dot{m} = - \gamma( 2 (Jm+h) ) \left( 
  1 + \mathrm{e}^{ -2 \beta (J m+h) }
 \right)
 \left( m - \tanh(\beta (J m + h )) \right).
\end{equation}
If $h=0$ the critical temperature is $T_\text{c} = J$, and above this temperature the equilibrium solution is $m^\text{eq} = 0$.
The inverse susceptibility is
\begin{equation}
 \chi^{\minus 1} =
 - J + T \equiv T - T_{\text{c}},
\end{equation}
therefore
\begin{equation}
\label{eq:lambda_min_mf}
 \lambda_{\text{min}} = \Gamma (T - T_{\text{c}}),
\end{equation}
where
$\Gamma = 2 \gamma(0; \beta) \beta = 2 \eta$ in the Ohmic bath.
Equation (\ref{eq:lambda_min_mf}) is the same result that we have already seen in figure \ref{fig:dynamic_crit_exponent}. As in the equilibrium statistical physics the mean field approximation gives back the exact result for the uniform model in the thermodynamic limit.

At the critical temperature the inverse susceptibility is zero, the linear term vanishes, and we need the higher order terms. Taylor expanding (\ref{eq:uniform_dot_m}) at $T=T_\text{c}$ around $m = m^\text{eq} \equiv 0$ up to third order gives.
\begin{equation}
 \delta \dot{m} = -\frac{ 2 }{ 3 } \eta J \delta m^3
\end{equation}
which has the
\begin{equation}
 \delta m(t) \propto t^{-\frac{ 1 }{ 2 }}
\end{equation}
solution for large $t$s, which means there isn't a characteristic time.

\begin{figure}
  \includegraphics[width=0.6\textwidth]{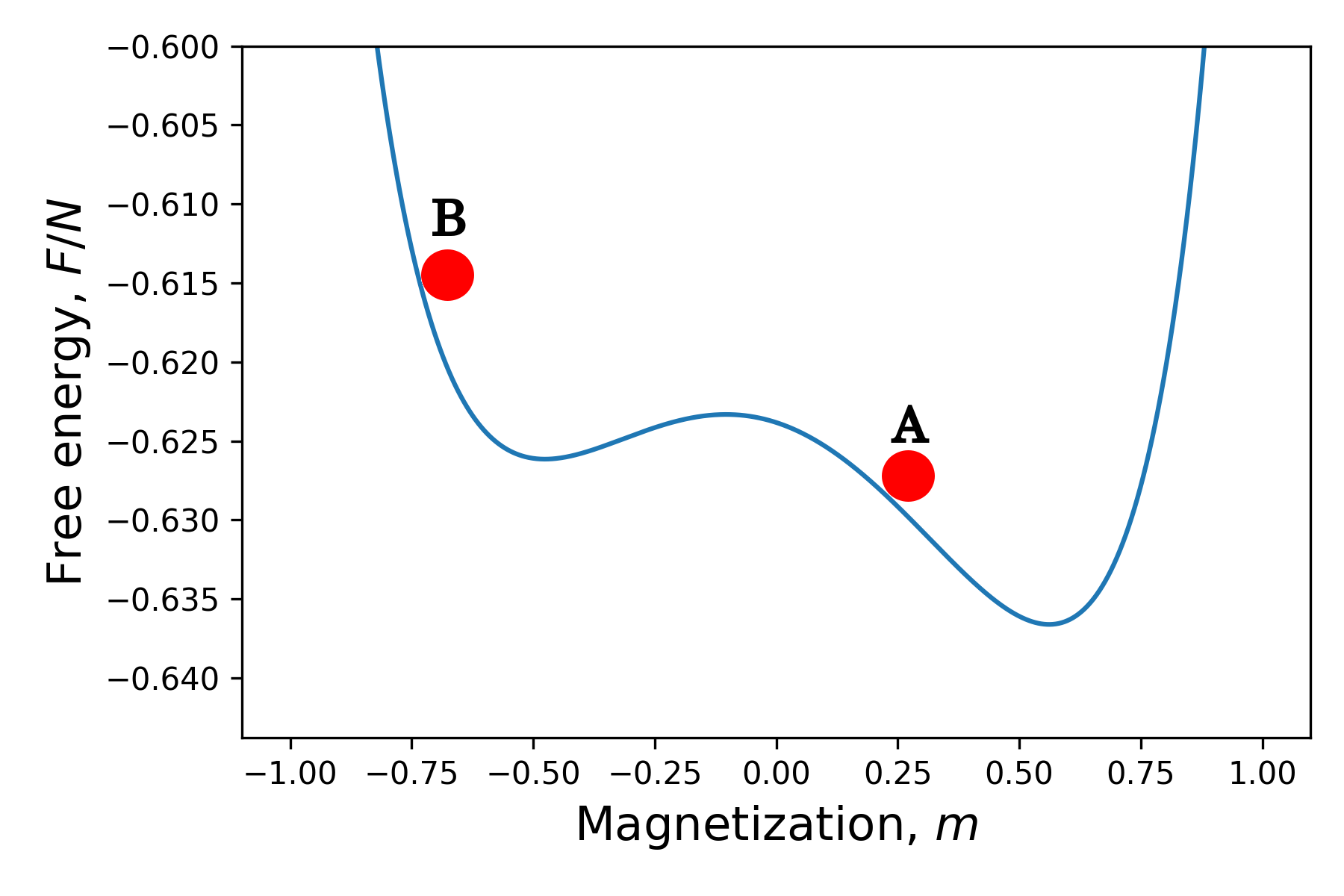}
 \caption{\textbf{Free energy of the uniform Ising model with finite external magnetic field}}
 \label{fig:uniform_free_energy}
\end{figure}

\begin{figure}
\centering
\captionsetup{justification=centering}
 \begin{subfigure}{0.45\textwidth}
  \includegraphics[width=\textwidth]{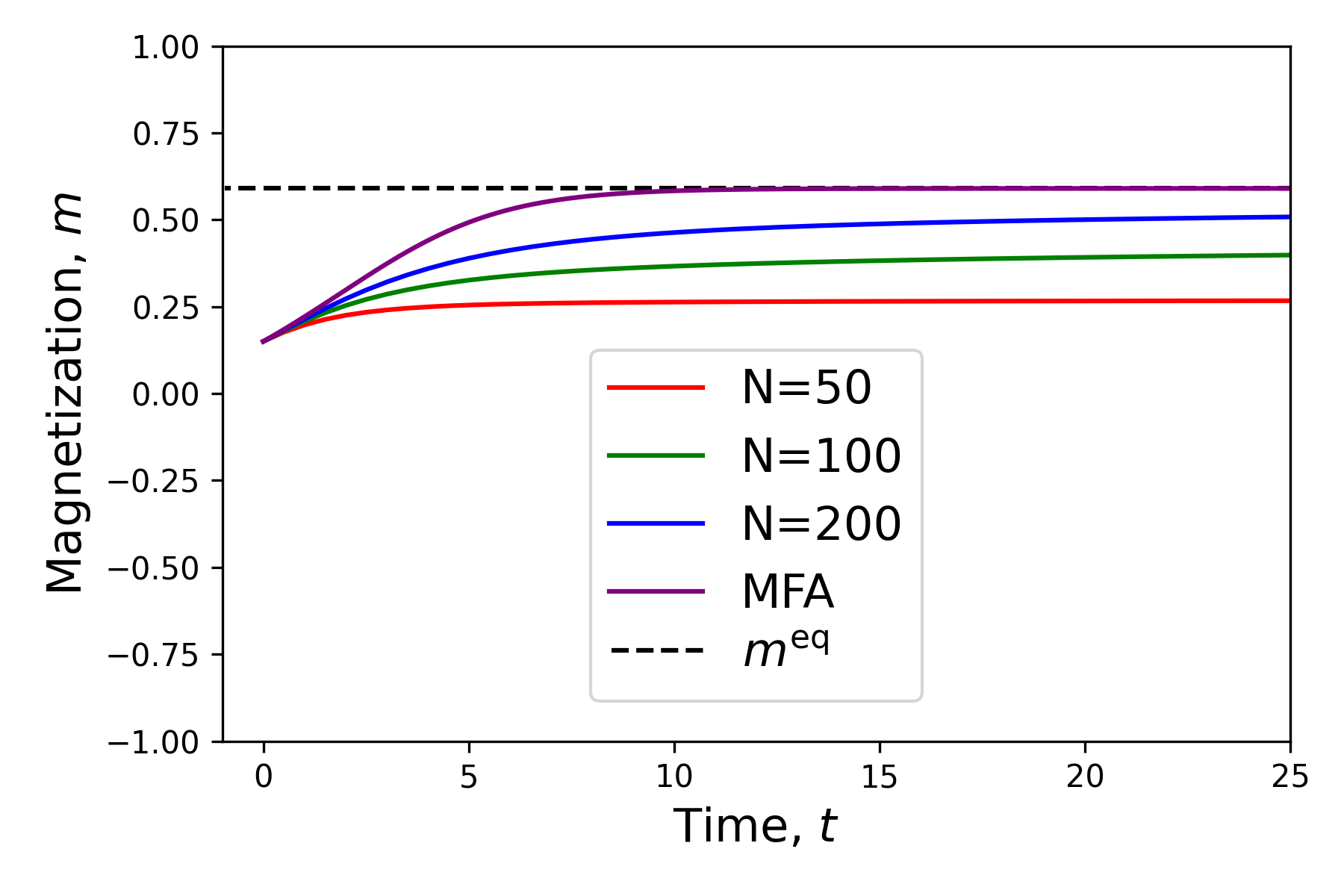}
  \caption{ \textbf{ Point A, Solutions when the initial $m$ is \ close to the global minimum } }
  \label{fig:point_A}
 \end{subfigure}
 \begin{subfigure}{0.45\textwidth}
  \includegraphics[width=\textwidth]{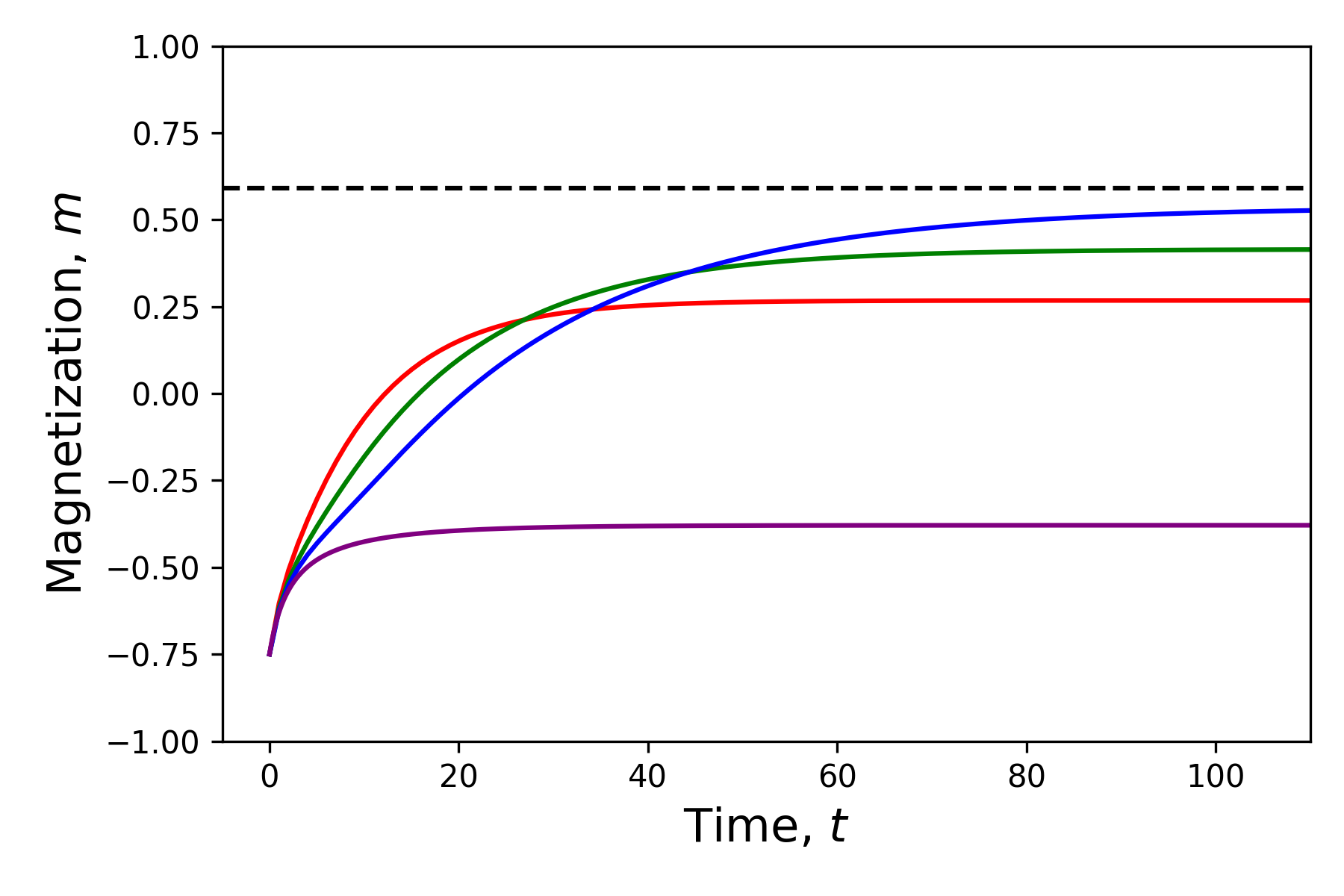}
  \caption{ \textbf{Point B, Solutions when the initial $m$ is close to the local minimum } }
  \label{fig:point_B}
 \end{subfigure}
 \caption{  
   \textbf{ Comparison of the time dependent mean field equation and the master equation }\\
  $J=1$, $\eta=1$, $h=0.02$, $T=0.9$, Ohmic bath
  }
\end{figure}

Below the critical temperature the linearized equation is good again, only the $m^\text{eq}$ changes. 
On the other hand equation (\ref{eq:uniform_dot_m}) can give different solution than the (\ref{eq:reduced_master_uniform2}) master equation.
If we want to compare these two equations the initial condition must be also the same which gives a restriction to the initial condition of (\ref{eq:reduced_master_uniform2}). In the mean field approximation the probability is a product of the one particle probabilities:
\begin{equation}
P_{\underline{S}}^\text{MFA} = \prod_{i} \frac{ 1+m_i S_i }{ 2 },
\end{equation}
which means the initial probability of (\ref{eq:reduced_master_uniform2}) is
\begin{equation}
 P_{N_\uparrow}(t=0;m) = \binom{N}{N_{\uparrow}} 
 \left( \frac{ 1 + m }{ 2 } \right)^{N_\uparrow}
 \left( \frac{ 1 - m }{ 2 } \right)^{N_\downarrow}
\end{equation}
If $h=0$, then the master equation must converge to the $m=0$ solution, but the time dependent mean field equation finds the global minimum of the free energy if initially $m \neq 0$.
If $h$ is finite and $m$ is in the valley of the global minimum (point A in figure \ref{fig:uniform_free_energy}), than in the $N \rightarrow \infty$ limit the solution of the master equation converges to the mean field solution (figure \ref{fig:point_A}).

On the other hand if initially $m$ is in the valley of the local minimum (point B in figure \ref{fig:uniform_free_energy}), then the solution  of the mean field equation converges to this local minimum, but the solutions of the master equation are totally different.
The probabilities converge to the equilibrium
\begin{equation}
P_{N_\uparrow}^\text{eq}(h,N) \propto
 \binom{N}{N_{\uparrow}} \mathrm{e}^{ - \beta E_{N_\uparrow}(h,N) }
\end{equation}
distribution, so they tend to approach $m^\text{eq}$ for large $N$ values in the $t \rightarrow \infty$ limit, but as $N$ grows, so does the relaxation time. In the thermodynamic limit the relaxation time diverges as in figure \ref{fig:N-lambda_large}. In the $N \rightarrow \infty$ limit the solution of the master equation converges to the solution of the mean field equation and none of them will approach the global minimum, because the relaxation time will be infinit.

\section{Conclusion}
In this work we have presented a Glauber-type master equation for the spin-boson model. Starting from the Redfield equation the population decoupled even without the secular approximation.
The most relevant dynamical properties are encoded in the  eigenvalues of the transition matrix of the master equation. They are temperature dependent, and behave significantly different below, above and at the critical temperature as a function of the system size.
In the case of the uniform, fully connected Ising model, in the thermodynamic limit, above the critical temperature the relaxation time follows a power law: $t_\text{r} \propto (T-T_\text{c})^{\minus 1} $.

We have derived a time dependent mean field equation, which is an effective equation of the master equation, containing only the $\langle S_i \rangle$ expected values. As every mean field theory it works best if the number of neighbours is large, so the fully connected Ising model is the best candidate, and the numerical simulations show that in the $N \rightarrow \infty$ limit the master equation gives back the same solutions as the mean field equation.

\section*{Acknowledgement}
This work was supported by NKFIH within the Quantum Technology
National Excellence Program (Project No. 2017-1.2.1-NKP-2017-00001) and
within the Quantum Information National Laboratory of Hungary, by the
ELTE Institutional Excellence Program (TKP2020-IKA-05) financed by the
Hungarian Ministry of Human Capacities, and Innovation Office (NKFIH)
through Grant No. K134437.

\bibliography{references.bib}
\end{document}